\documentclass[twocolumn,showpacs]{revtex4}
\usepackage[xdvi]{graphicx}%
\usepackage{dcolumn}
\usepackage{amsmath}
\makeatletter

\def\btt#1{\texttt{\@backslashchar#1}}%
\DeclareRobustCommand\bblash{\btt{\@backslashchar}}%
\makeatother

\begin{document}

\title{Memory effect on the formation of drying cracks}
\author{Michio Otsuki }
\email{otsuki@jiro.c.u-tokyo.ac.jp}

\affiliation
{Department of Pure and Applied Sciences,  
University of Tokyo, Komaba, Tokyo 153-8902, Japan}

\date{\today}

\begin{abstract}
We propose a model for the formation of drying cracks in a viscoplastic material.
In this model, 
we observe that when an external force is applied to a viscoplastic material
before drying,
the material memorizes the effect of the force as a plastic deformation. 
The formation of the drying cracks is influenced by this plastic deformation.
This outcome clarifies the result of a recent experiments which demonstrated 
that a drying fracture pattern on a powder-water mixture 
depends on the manner in which an external force is applied before drying.
We analytically express the position of the first crack 
as a function of the strength of an external force
applied before drying.
From the expression,
we predict that there exists a threshold on the strength of the force.
When the force applied is smaller than the threshold,
the first crack is formed at the center of the mixture;
however, when the force applied exceeds the threshold,
the position of the first crack deviates from the center.
The extent of the deviation increases 
as a linear function of the difference between 
the strength of the force and the threshold.
\end{abstract}

\maketitle

\section{INTRODUCTION}

Cracks are observed on various materials such as 
rocks, tectonic plates, and paintings.  These cracks are 
fascinating and has been studied by many researchers.
The study of drying cracks in a powder-water mixture
was also included in these researchers.
In one example, 
it was noticed that when a layer 
of a powder-water mixture  is dried in a container, 
it shrinks and cracks are formed on it \cite{groisman}.
The cracks extend from the surface of the mixture to the bottom 
and propagate horizontally along a line. As a result, 
a two-dimensional fracture pattern is formed on the surface of the mixture.

When we gently pour a powder-water mixture into a container and 
leave it undisturbed during the drying process, a random, isotropic 
fracture pattern is formed. However,  
Nakahara and Matsuo reported that when an external force is applied to 
the mixture before drying, the fracture pattern changes 
depending on the manner in which the force is applied \cite{nakahara}.
For example, when the mixture is vibrated in one direction before drying, 
cracks that are perpendicular to the direction of the vibration 
emerge first.
Finally, a lamellar fracture pattern is formed. 
It takes more than 3 days for the cracks to be formed after the vibration.
It is quite surprising that the effect of applying the force remains 
for such a long time.

The experimental result attained by Nakahara and Matsuo 
represents that a fracture pattern is
controlled by the memory of an external force applied before drying. 
Similar memory effects that a response 
can be controlled by the memory of an operation have been observed
in other materials,
such as sand piles \cite{sand}, micro-gel pastes \cite{paste},
and rubbers \cite{rubber}.  
By recalling that 
these memory effects have been studied from a rheological point of view,
we conjecture that the rheological property of a mixture 
plays an important role in the memory 
effect on the formation of drying cracks.

Among the rheological properties of the mixture, the most conspicuous 
one may be plasticity. Hence, we study the role of plasticity 
in the memory effect on the formation of drying cracks. 
First, in Sec. \ref{modelsec},
we propose 
a model of the formation of drying cracks in a viscoplastic material.
In Sec. \ref{asec}, we find that 
a viscoplastic material memorizes 
the effect of an external force before drying 
as a plastic deformation by calculating the model numerically. 
By the influence of this plastic deformation,
when an external force is applied to the material, 
a crack perpendicular to the force emerges earlier
than when no force is applied.
Based on this result, we conjecture  
that the perpendicular cracks emerge first 
by a plastic deformation.
Furthermore, we express
the position of the first perpendicular crack in terms of measurable material 
properties in Sec. \ref{zuresec}. 
This result can be used to test 
our conjecture that the memory effect is caused by a plastic deformation.
Section \ref{sumsec} is devoted to the summary and discussion.
Technical details are summarized in Appendix \ref{sol}.

\section{MODEL}
\label{modelsec}

\begin{figure}[htbp]
    \begin{center}
    \includegraphics[height=10em]{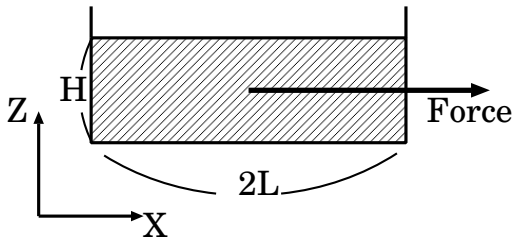}
    \caption{The illustration of viscoplastic material in a container.}
    \label{paste2}
    \end{center}
    \end{figure}

We propose the model of a viscoplastic material in a similar way as 
that demonstrated by Ooshida and Sekimoto \cite{waribasi}. 
We consider 
a viscoplastic material of thickness $H$ and width $2L$ in a container, 
as shown in Fig. \ref{paste2}. The coordinate system $(x,z)$ is 
assumed  such that the center of the container is at $x=0$ and 
the bottom is at $z=0$. For the mathematical 
simplicity, we restrict our attention to plane strain deformations 
of the viscoplastic material and we consider only a displacement 
$u(x,z,t)$ in the $x$ direction and a plastic strain $s(x,z,t)$,
which express the occurrence of a plastic deformation.
We assume  that the time evolutions of $u(x,z,t)$ and  $s(x,z,t)$ are described as 
    \begin{eqnarray}
      \gamma \frac{\partial u(x,z,t)}{\partial t} 
      = \frac{\partial \sigma_{xx}(x,z,t)}{\partial x} +
      \frac{\partial \sigma_{xz}(x,z,t)}{\partial z} + \alpha(t) ,\label{u}  
    \end{eqnarray}
    \begin{eqnarray}
      B \frac{\partial s(x,z,t)}{\partial t} =
      \left \{
      \begin{array}{cc}
	0 & |\sigma_{xz}|<\sigma_Y(t) \\
	(|\sigma_{xz}| - \sigma_Y(t))
	 \displaystyle \frac{\sigma_{xz}}{|\sigma_{xz}|} & {\rm otherwise},
      \end{array}
      \right . \label{s}
    \end{eqnarray}
where $\gamma$ is a coefficient of a viscosity, 
$\alpha(t)$ is an external force, $\sigma _Y(t)$ is a yield stress,
$\sigma_{xx}(x,z,t)$ is a normal stress, 
$\sigma_{xz}(x,z,t)$ is a shear stress
and $B$ is a coefficient which determines the speed 
of the plastic strain.
$\sigma_{xx}(x,z,t)$ and $\sigma_{xz}(x,z,t)$ are determined
by constitutive equations: 
    \begin{eqnarray}
      \sigma_{xx}(x,z,t) = (\lambda + 2\mu)
      \left( \frac{\partial u(x,z,t)}{\partial x}+c(t) \right ) \label{sxx}, \\
      \sigma_{xz}(x,z,t) =  \mu\left(\frac{\partial u(x,z,t)}{\partial z} - s(x,z,t)\right) ,\label{sxz} 
    \end{eqnarray}
where $\lambda$ and $\mu$ are lame coefficients, and  $c(t)$ is a 
reference strain rate.
An increase in the reference strain rate represents an effect 
of shrinking by drying \cite{kitsune,sasa}. 

We consider the case
where the functional forms of $\alpha(t)$, $\sigma_Y(t)$, and $c(t)$
are given as
\begin{eqnarray}
       \alpha(t) =
      \left \{
      \begin{array}{cc}
	\alpha_{M} \sin(\pi t /T_1) & t < T_1 \\
	0 &  t > T_1,
      \end{array}
      \right.
      \label{alpha}
\end{eqnarray}
\begin{eqnarray}
       \sigma_Y (t) =
      \left \{
      \begin{array}{cc}
	\sigma _{Y0} & t < T_2 \\
	\infty & t >T_2,
      \end{array}
      \right. 
      \label{sY}
    \end{eqnarray}
\begin{eqnarray}
       c(t) =
      \left \{
      \begin{array}{cc}
	0 & t < T_3 \\
	b (t- T_3) & t>T_3,
      \end{array}
      \right. 
      \label{c}
\end{eqnarray}
where 
$\alpha_M$, $\sigma_{Y0}$, and $b$ represent
the maximum value of an external force,
the yield stress before drying,
and the speed of drying, respectively. 

In order to specify the model completely,
we assume the boundary conditions. 
At the region where the material comes into contact with the container, 
the displacement is set to zero. 
In contrast, 
at the region where there is no contact between the material and the container, 
the stresses applied to the free surface become zero.  
Here, one peculiar phenomenon arises:
when the material is dried,
it peels off from the walls of the container.
This phenomenon implies that the boundary conditions change when
the peeling occurs. 
Hence, we assume boundary conditions
    \begin{eqnarray}
      u(\pm L,z,t)=0,\nonumber \\
      u(x,0,t)=0,  \label{bbc} \\
      \sigma_{xz}(x,H,t)=0,\nonumber
    \end{eqnarray}
for $0<t<T_2$,
and
    \begin{eqnarray}
      \sigma_{xx}(\pm L,z,t)=0,\nonumber \\
      u(x,0,t)=0,  \label{abc} \\
      \sigma_{xz}(x,H,t)=0,\nonumber
    \end{eqnarray}
for $t>T_2$, 
where we consider that the peeling occurs at $t=T_2$. 
In Table \ref{paratable},  we summarize the functional forms of the parameters
and the boundary conditions at the walls of the container
when $T_1<T_2<T_3$. 

\begin{table*}[htbp]
\begin{ruledtabular}
\caption{\label{paratable} Boundary conditions (B.C.) and parameters}
	\begin{tabular}{c|ccccc}
	   & $t$ & $\alpha$ & $c$ & $\sigma _Y$ 
	  & B.C. at walls\\ \hline
	   Applying an external force & $0<t<T_1$ & $\alpha_{M} \sin(\pi t /T_1)$ & $0$ & $\sigma _{Y0}$ & 
	  $u(\pm L,z,t)=0$ \\ \hline
	   Relaxation & $T_1<t<T_2$ & $0$ & $0$ & $\sigma _{Y0}$ 
	  & $u(\pm L,z,t)=0$ \\ \cline{2-6}
	   & $T_2<t<T_3$ & $0$ & $0$ & $\infty$ & $\sigma_{xx}(\pm L,z,t)=0$ \\ \hline
	   Drying & $T_3<t $ & $0$ & $b(t-T_3)$ & $\infty$ 
	  & $\sigma_{xx}(\pm L,z,t)=0$ 
	\end{tabular} 
\end{ruledtabular}
\end{table*}

Finally, we 
assume the condition of a crack formation.  
To the best of our knowledge, the 
condition has not yet been completely understood; however, two conditions 
have been used in previous works.  These are the critical stress 
condition and the Griffith criterion. Under the critical 
stress condition, a crack is formed when the stress 
exceeds a material constant.  This has been used in several numerical 
models \cite{kitsune} because of the technical  advantage of 
the local condition. In contrast, under the Griffith criterion,
a crack is formed when the energy released during the formation of 
a crack exceeds the increase of the surface energy.
This condition has been used by Komatsu and Sasa 
in their theory \cite{sasa}. Although we cannot determine which 
condition is more efficient,  we employ the critical stress condition for 
the simplicity of the treatment.  Concretely, we define an 
average normal stress at the position $x$ as 
    \begin{eqnarray}
      \left < \sigma_{xx} (x,t) \right >=\frac{1}{H}
      \int dz \ \sigma _{xx} (x,z,t).
    \end{eqnarray}
Then, the condition of a crack formation is given as 
    \begin{eqnarray}
      \left < \sigma_{xx} (x_c,t_c) \right > =\sigma _b \rightarrow 
      \textrm{$\sigma _{xx}(x_c,z,t) = 0$  for $t>t_c$},
      \label{fc}
    \end{eqnarray}
where $x_c$ is the position of a crack, 
and $t_c$ is the time when the crack is formed. 
We only consider cracks that are perpendicular to the $x$ axis.

To summarize,  our model consists of  Eqs. (\ref{u}) and (\ref{s})
with parameters given by Eqs. (\ref{alpha}), (\ref{sY}), and (\ref{c})
under the boundary conditions (\ref{bbc}) and (\ref{abc}), 
and the condition of a crack formation (\ref{fc}).

\section{A QUALITATIVE COMPREHENSION OF THE ORIGIN OF THE MEMORY EFFECT}
\label{asec}

    In order to understand how a viscoplastic material 
    memorizes the influence of an external force that is applied before drying, 
    we numerically calculate Eqs. (\ref{u}) and (\ref{s})
    with the control parameter $\alpha_M$.
    In this numerical calculation, the initial conditions are
    given as $u(x,z,0)=s(x,z,0)=0$, 
    and the parameter values are set as $H=1.0$, $L=10.0$,
    $\lambda=1.0$, $\mu=0.1$, $\sigma_{Y0}=0.05$, $\sigma_b=0.01$,
    $\gamma = 1.0$, $ B=1.0$, $T_1=30$, $T_2 = 60$, $T_3 = 90$, and 
    $b=0.00019$.
    In the following we report all quantities in reduced units, i.e., 
    length in units of $H$, stress in units of $\lambda$
    , time in units of $(\gamma H^2/\lambda)$
    and other quantities in units of the combinations of these units.

    When no external force is applied to the material 
    before drying, 
    (that is, $\alpha_M=0$)
    stresses become zero at $t=T_3$ (just before starting drying process).  
    Even when an external force is applied to a material,
    provided  $\alpha_M$ is not so large 
    that the shear stress exceeds the yield stress,
    the stresses remain zero at $t=T_3$. 
    However, when $\alpha_M$ is sufficiently large, 
    the stresses have a non-zero value at $t=T_3$, 
    by a plastic deformation.

\begin{figure}[htbp]
    \begin{center}
    \includegraphics[height=15em]{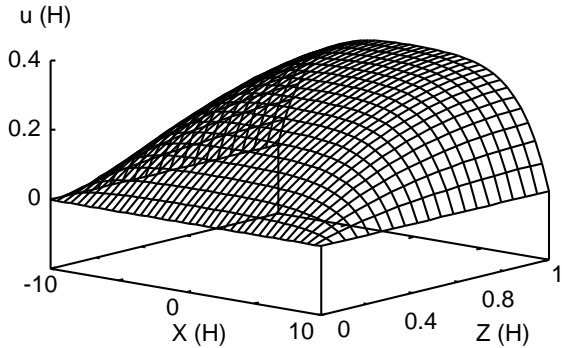}
    \caption{
      $u(x,z,T_1/2)$ 
      as a function of $(x,z)$
      in the case $\alpha_M=0.08$. 
    }
    \label{A2u1}
    \end{center}
    \end{figure}

\begin{figure}[htbp]
    \begin{center}
    \includegraphics[height=15em]{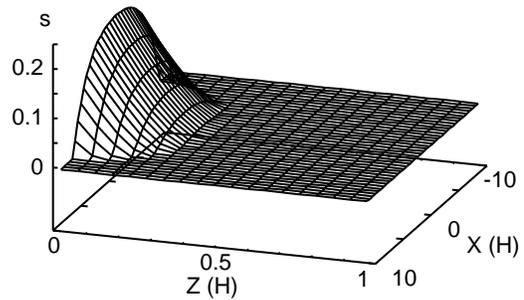}
    \caption{
      $s(x,z,T_3)$ 
      as a function of $(x,z)$
      in the case $\alpha_M=0.08$. 
    }
    \label{A2s2}
    \end{center}
    \end{figure}

\begin{figure}[htbp]
    \begin{center}
    \includegraphics[height=15em]{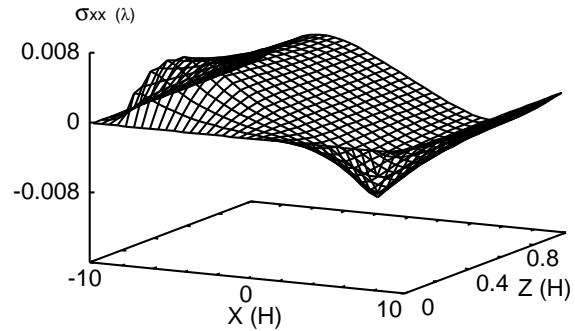}
    \caption{
      $\sigma_{xx}(x,z,T_3)$ 
      as a function of $(x,z)$
      in the case $\alpha_M=0.08$. 
    }
    \label{A17tauxL2s3}
    \end{center}
    \end{figure}

    As an  example, we show the results in the case $\alpha_M =0.08$.
    When an external force $\alpha(t)$ increases in time from 0,
    the material displaces to the direction of $x$, 
    as shown in Fig. \ref{A2u1}.
    Due to the influence of the boundary conditions $u(\pm L,z,t)=0$,
    the material is pulled in the left region ($x<0$) 
    and pushed in the right region ($x>0$).
    When the deformation becomes sufficiently large, 
    the shear stress $\sigma_{xz}(x,z,t)$ 
    exceeds the yield stress $\sigma_{Y0}$. 
    Then, a plastic deformation occurs near the bottom ($z=0$),
    as shown in Fig. \ref{A2s2}.
    Due to this plastic deformation, 
    the material remains pulled in the left region ($x<0$) 
    and pushed in the right region ($x>0$) 
    even after an external force $\alpha(t)$ becomes 0.
    Therefore, the normal 
    stress $\sigma_{xx}(x,z,T_3)$ is positive in the left region ($x<0$) 
    and negative in the right region ($x>0$),
    as shown in Fig. \ref{A17tauxL2s3}.
    In this manner, 
    a material memorizes the effect of the force as a plastic deformation
    when an external force is applied to the material before drying.

\begin{figure}[htbp]
    \begin{center}
    \includegraphics[height=15em]{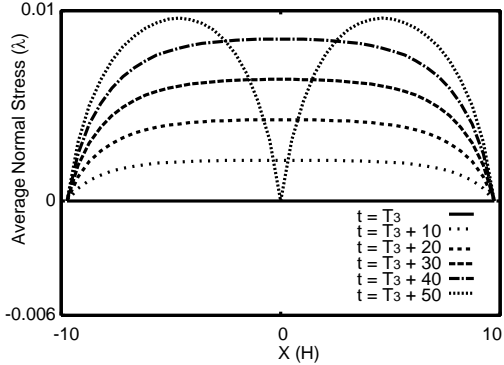}
    \caption{
    The average normal stress $\left <\sigma_{xx} (x,t) \right>$ 
    in the case $\alpha_{M}=0$
    as a function of $x$. 
    }
    \label{A0tauxL1L2}
    \end{center}
    \end{figure}

\begin{figure}[htbp]
    \begin{center}
    \includegraphics[height=15em]{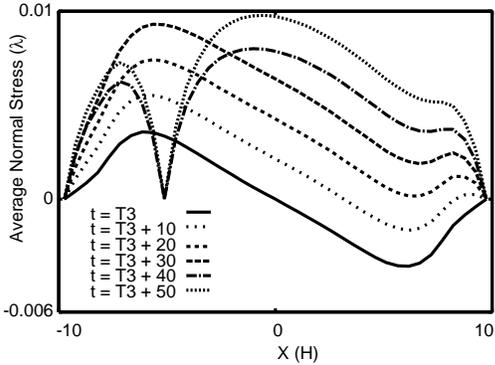}
    \caption{
    The average normal stress $\left <\sigma_{xx} (x,t) \right>$ 
    in the case $\alpha_{M}=0.08$
    as a function of $x$. 
    }
    \label{A2tauxL1L2}
    \end{center}
    \end{figure}

  Next, 
  we investigate the influence of a plastic deformation 
  on the formation of drying cracks
  by comparing the time evolutions of $\left <\sigma_{xx}(x,t) \right >$ 
  after starting the drying process ($t\geq T_3$)
  in the cases 
  $\alpha_M=0$ and $\alpha_M=0.08$, 
  which are illustrated 
  in Fig. \ref{A0tauxL1L2} and Fig. \ref{A2tauxL1L2}.
  In the case $\alpha_M=0$, $\left <\sigma_{xx}(x,t) \right >=0$
  just before drying ($t=T_3$);
  however, in the case $\alpha_M=0.08$,
  $\left <\sigma_{xx}(x,t) \right>$ is positive in the left region ($x<0$) 
  and negative in the right region ($x>0$)
  by a plastic deformation.
  As the material is dried, the average normal stress 
  $\left <\sigma_{xx}(x,t) \right>$ increases
  in a similar manner for both cases.
  However, until a crack is formed, 
  the maximum values of $\left <\sigma_{xx}(x,t) \right>$ 
  in the case $\alpha_M=0.08$
  are larger than those in the case $\alpha_M=0$.
  Due to this large stress,
  a crack is formed earlier in the case $\alpha_M=0.08$, 
  than in the case $\alpha_{M}=0$. 

    Here, we consider the correspondence between the results of 
    the viscoplastic model and the experiments.
    In the viscoplastic model, 
    we consider a crack that is perpendicular to an external force 
    applied before drying.
    Due to the influence of a plastic deformation,
    the crack is formed earlier
    when an external force is applied to the material 
    than when no force is applied. 
    Therefore, 
    we conjecture that in the experiments too, 
    the perpendicular crack emerges 
    due to the influence of a plastic deformation
    when an external force is applied to the material.

\section{QUANTITATIVE PREDICTIONS}
\label{zuresec}

    In the viscoplastic model,
    the position of the first perpendicular crack deviates
    from the center of the material 
    to the opposite direction of an external force applied before drying
    as shown in Fig. \ref{A2tauxL1L2}. 
    We wish to analytically express 
    the position of the first perpendicular crack $x_c$ 
    as a function of the maximum value $\alpha_M$ of the external force. 
    However, since it is difficult to study 
    the partial differential equations (\ref{u}) and (\ref{s}),
    we simplify the equations on some assumptions.

    First, we assume that an external force $\alpha(t)$ and 
    a reference strain rate $c(t)$ vary more slowly 
    than the relaxation of 
    a displacement $u(x,z,t)$ and a plastic strain $s(x,z,t)$.
    Then, it is sufficient to calculate the static 
    solution of the equations, given an external force $\alpha(t)$, 
    a yield stress $\sigma_{Y}(t)$, and a reference strain rate $c(t)$.
    Second, we discretize the partial derivative of $z$ as 
    \begin{eqnarray}
    \frac{\partial u(x,H,t)}{\partial z} = \frac{u(x,H,t)-u(x,0,t)}{H} 
    = \frac{u(x,H,t)}{H},
    \end{eqnarray}
    \begin{equation}
    \frac{\partial \sigma_z(x,H,t)}{\partial z}  = 
    \frac{\sigma_{xz}(x,H,t)-\sigma_{xz}(x,0,t)}{H} 
     =  \frac{-\sigma_{xz}(x,0,t)}{H},
    \end{equation}
    where $u(x,0,t) = 0$ and $\sigma_{xz}(x,H,t) = 0$
    (see Eqs. (\ref{bbc}) and (\ref{abc})).
    Based on these assumptions, the displacement at the surface
    $U(x,t)=u(x,H,t)$,  
    the normal stress at the surface $T_{xx}(x,t)=\sigma_{xx}(x,H,t)$, 
    and the shear stress at the bottom $T_{xz}(x,t)=\sigma_{xz}(x,0,t)$ 
    are determined by the equations
    \begin{eqnarray}
    \frac{\partial T_{xx}(x,t)}{\partial x} -\frac{T_{xz}(x,t)}{H} + \alpha(t) =0,\label{dut}
    \end{eqnarray}
    \begin{eqnarray}
      T_{xx}(x,t) = (\lambda + 2\mu)
      \left(\frac{\partial U(x,t)}{\partial x} + c(t)\right),\label{dsxx} \\
      T_{xz}(x,t) =  \mu\left ( \frac{U(x,t)}{H} - S(x,t)\right ), \label{dsxz}
    \end{eqnarray}
    where $S(x,t)$ is a plastic strain at the surface ($s(x,H,t)$).
    The time evolution of $S(x,t)$ is as follows:
    if 
    \begin{eqnarray}
    |T_{xz}(x,t)|<\sigma_Y(t), 
    \end{eqnarray}
    $S(x,t)$ does not change in time.
    However,
    if
    \begin{eqnarray}
    |T_{xz}(x,t)|>\sigma_Y(t), 
    \end{eqnarray}
    $S(x,t)$ is determined by the condition
    \begin{eqnarray}
    |T_{xz}(x,t)|=\sigma_Y(t), 
    \end{eqnarray}
    which yields 
    \begin{eqnarray}
    S(x,t) = \frac{U(x,t)}{H} \pm \frac{\sigma_Y(t)}{\mu}. \label{S}
    \end{eqnarray}
    Here, the sign depends on the sign of $T_{xz}(x,t)$.
    By substituting Eqs. (\ref{dsxx}) and (\ref{dsxz}) into Eq. (\ref{dut}),
    we obtain the equation of $U(x,t)$ as 
    \begin{eqnarray}
    (\lambda + 2\mu)\frac{\partial^2 U}{\partial x^2} 
    -\frac{\mu}{H^2}U +\frac{\mu}{H}S+ \alpha(t) =0.\label{du}
    \end{eqnarray}

    The functional forms of $\alpha(t)$, $\sigma_Y(t)$, and $c(t)$
    are given as Eqs. (\ref{alpha}), (\ref{sY}), and (\ref{c}), respectively.
    From Eqs. (\ref{bbc}) and (\ref{abc}), the boundary conditions are 
    rewritten as 
    \begin{eqnarray}
    U(L,t)=U(-L,t)=0 \label{dbbc}
    \end{eqnarray}
    for $0<t<T_2$ and
    \begin{eqnarray}
    T_{xx}(L,t)=T_{xx}(-L,t)=0 \label{dabc}
    \end{eqnarray}
    for $T_2<t$.

    From Eq. (\ref{fc}), 
    the condition of a crack formation is rewritten as
    \begin{eqnarray}
    T_{xx}(x_c,t_c)=\sigma_b. \label{txc}
    \end{eqnarray}
    Moreover, because $T_{xx}(x,t_c)$ has a maximum value at $x=x_c$,
    the equation
    \begin{eqnarray}
    \frac{\partial T_{xx}(x_c,t_c)}{\partial x}=0 \label{dtxc}
    \end{eqnarray}
    should be satisfied.

    From these equations, $x_c$ can be calculated.
    The result is summarized below 
    (see Appendix \ref{sol} for details of 
    the calculation).
    First, we denote the threshold value of 
    the external force by $\alpha_{Y0}$,
    which is derived as
    \begin{eqnarray}
    \alpha_{Y0} = \frac{\sigma_{Y0} \cosh qL}{H(\cosh qL-1)},\label{as}
    \end{eqnarray}
    where
    \begin{eqnarray}
    q=\sqrt{ \frac{\mu}{(\lambda + 2\mu)H^2}}.
    \end{eqnarray}
    Then, if $\alpha_M<\alpha_{Y0}$,
    the position of the first perpendicular crack $x_c$ is expressed as
    \begin{eqnarray}
    x_c=0. \label{xc0}
    \end{eqnarray}

    In contrast, when $\alpha_{Y0}<\alpha_M<\alpha_{Y1}$,
    $x_c$ is determined by 
    \begin{eqnarray}
	  \cosh qL  & = &
	   B(0,x_c) + \frac{\sigma_b qH}{\sigma_{Y0} -\alpha_{M}H} \sinh qx_c \nonumber \\
	  & & -\frac{D(L,x_s)}{\sinh qL}(1-\cosh qx_c \cosh qL), \label{xc}
    \end{eqnarray}
    where $\alpha_{Y1}$ is defined as
    \begin{eqnarray}
    \alpha_{Y1}=\frac{\sigma_{Y0} (2\cosh qL - B(L,x_s))}{\cosh qL -B(L,x_s)}.
    \end{eqnarray}
    Here, $B(x_1,x_2)$ and $D(x_1,x_2)$ are defined as
    \begin{eqnarray}
    B(x_1,x_2) = \cosh q(x_1-x_2) + qx_2 \sinh(x_1-x_2), 
    \end{eqnarray}
    \begin{eqnarray}
    D(x_1,x_2) = \sinh q(x_1-x_2) + q x_2\cosh q(x_1-x_2).
    \end{eqnarray}
    $x_s$ represents the region where a plastic deformation occurs
    as $-x_s<x<x_s$,
    which is determined by
    \begin{eqnarray}
    \alpha_M H + (\sigma_{Y0}-\alpha_M H) B(L,x_s)= 0. \label{xs}
    \end{eqnarray}

    We further extract a simple expression of $x_c$ by 
    focusing on the region where $\alpha_M$ is adjacent to $\alpha_{Y0}$.
    Assuming that
    \begin{eqnarray}
	    qx_c \ll 1, 
    \end{eqnarray}
    \begin{eqnarray}
	    qx_s \ll 1, 
    \end{eqnarray}
    and expanding Eqs. (\ref{xc}) and (\ref{xs}) to the first order of $x_c$ 
    and to the second order of $x_s$,
    we obtain 
    \begin{eqnarray}
	    x_c=-\frac{(\cosh qL -1)^2}{q^2\sigma_b \cosh qL} 
	    (\alpha_M - \alpha_{Y0}).
	    \label{xceq}
    \end{eqnarray}

\begin{figure}[htbp]
    \begin{center}
    \includegraphics[height=15em]{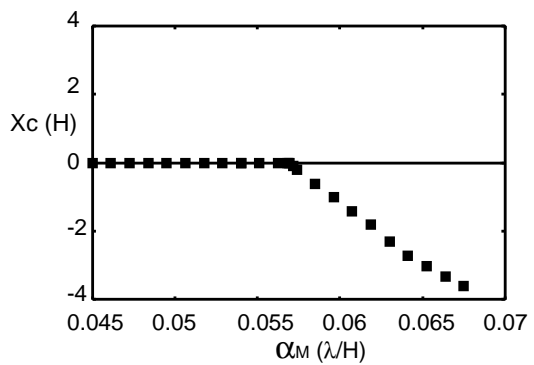}
    \caption{
    The position of the first perpendicular crack $x_c$ 
    as a function of the maximum 
    value of an external force $\alpha_{M}$
    for parameter set A. 
    }
    \label{data}
    \end{center}
    \end{figure}

\begin{figure}[htbp]
    \begin{center}
    \includegraphics[height=15em]{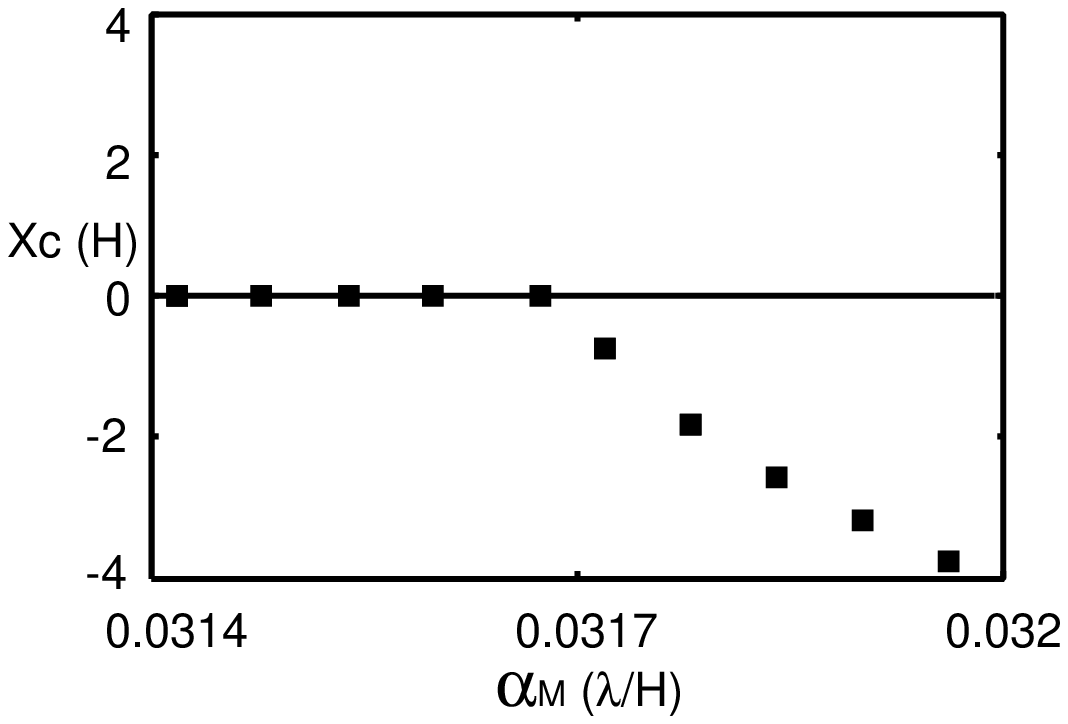}
    \caption{
    The position of the first perpendicular crack $x_c$ 
    as a function of the maximum 
    value of an external force $\alpha_{M}$
    for parameter set B. 
    }
    \label{datb}
    \end{center}
    \end{figure}

\begin{figure}[htbp]
    \begin{center}
    \includegraphics[height=15em]{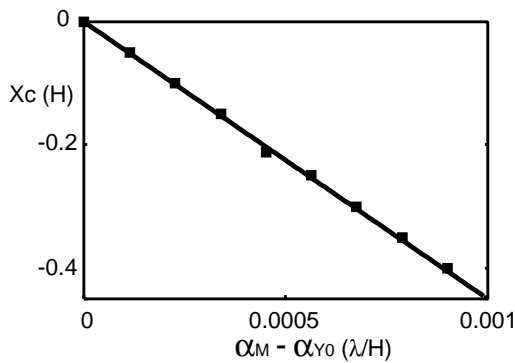}
    \caption{
    The position of the first perpendicular crack $x_c$ 
    as a function of difference between the maximum 
    value of an external force $\alpha_{M}$
    and the threshold value $\alpha_{Y0}$
    for parameter set A. 
    }
    \label{datd}
    \end{center}
    \end{figure}

    In the calculation of $x_c$,
    we simplified the original partial differential equations
    based on some assumptions. 
    In order to confirm 
    the qualitative accuracy of Eqs. (\ref{xc0}) and (\ref{xceq}),
    we numerically calculate the solutions of $x_c$
    for the original partial differential equations.
    In Fig. \ref{data} and \ref{datb},
    we show the numerical solutions of $x_c$
    for the case 
    where the parameter values are same as in Sec. \ref{asec} 
    (parameter set A)
    and for the case where the parameter values are set as 
    $H=1.0$, $L=15.0$,
    $\lambda=1.0$, $\mu=0.2$, $\sigma_{Y0}=0.03$, $\sigma_b=0.02$,
    $\gamma = 1.0$, $ B=1.0$, $T_1=300$, $T_2 = 600$, $T_3 = 500$, and 
    $b=0.000005$ (parameter set B).
    There exists a threshold value $\alpha_{Y0}$.
    $x_c$ remains zero when the external force $\alpha_M$
    is smaller than the threshold value $\alpha_{Y0}$
    for both of the parameter sets.
    When the external force $\alpha_M$
    is larger than the threshold value $\alpha_{Y0}$, $x_c$ deviates from zero.
    In Fig. \ref{datd},
    we show the numerical solution of $x_c$
    as a function of difference between the maximum 
    value of an external force $\alpha_{M}$
    and the threshold value $\alpha_{Y0}$
    for parameter set A. 
    This figure clearly indicates 
    $x_c$ deviates from zero as a linear function of 
    the difference between $\alpha_M$ and $\alpha_{Y0}$
    when $\alpha_M$ is adjacent to $\alpha_{Y0}$.
    These behavior qualitatively 
    agree with Eqs. (\ref{xc0}) and (\ref{xceq}) 
    though the values of $\alpha_{Y0}$ are different.

    Based on these results, we expect that 
    in experiments too,
    the relation between the position of the first crack
    and the maximum value of an external force is expressed as 
    Eqs. (\ref{xc0}) and (\ref{xceq}).
    The experimental confirmation of Eqs. (\ref{xc0}) and (\ref{xceq})
    supports our conjecture that 
    the memory effect on the formation of drying cracks 
    arises from a plastic deformation of the material.

\section{SUMMARY AND DISCUSSION}
\label{sumsec}

  In this paper, 
  we model the formation of drying cracks in a viscoplastic material.
  In numerical experiments, we observe that 
  when an external force is applied before drying, 
  a crack whose direction is perpendicular to the force
  emerges earlier than when no force is applied.
  This phenomenon occurs 
  because of a plastic deformation.
  Based on this observation, 
  we conjecture that a plastic deformation
  is the cause of the memory effect on the formation of drying cracks.
  In order to check this theory,
  we quantitatively predict the position 
  of the first perpendicular crack.
  If Eqs. (\ref{xc0}) and (\ref{xceq}) are experimentally confirmed, 
  it may be concluded 
  that a plastic deformation causes the 
  memory effect on the formation of drying cracks.

  Here, we remark on the neglect of some quantities in our model.
  A displacement $w$ in the z-direction is not taken into account in our model.
  Moreover, normal stresses $\sigma_{xx}$, $\sigma_{yy}$ and $\sigma_{zz}$ are 
  not taken into account in the equation 
  which represents the occurrence of a plastic deformation, i.e.,
  Eq. (\ref{s}).
  The reason for neglecting $w$ is that
  $w$ plays little role in the formation of the crack
  because a crack is caused by a normal stress $\sigma_{xx}$
  and $w$ hardly contributes to $\sigma_{xx}$.
  The reason for neglecting 
  $\sigma_{xx}$, $\sigma_{yy}$ and $\sigma_{zz}$ is that
  we consider the situation where a plastic deformation occurs by a shear.
  To make sure that the neglect of these quantities does not affect our results,
  we simulated the model in which these quantities were taken into account.
  As far as we checked,
  we obtained qualitatively similar results with the model in present paper,
  for example,
  with regard to space distributions of a horizontal displacement $u(x,z,t)$, 
  stresses $\sigma_{xx}(x,z,t)$ and $\sigma_{xz}(x,z,t)$ 
  after applying an external force.

  In addition, we remark on the the results 
  for another choice of the parameters and the expressions 
  of $\alpha(t)$, $\sigma_Y (t)$ and $c(t)$.
  As far as we checked,
  we observe the same phenomenon 
  that the first perpendicular crack occurs earlier 
  and the position of the crack deviates from center
  for any choice of the parameter
  as shown in Fig. \ref{data} and Fig. \ref{datb}. 
  For any choice of the expressions
  of $\alpha(t)$, $\sigma_Y (t)$ and $c(t)$,
  the same phenomenon are observed. 
  Hence, we expect that our results are robust 
  for the variation of the parameters and the expressions.

  Recently, Nakahara and Matsuo measured  
  the rheological property of a powder-water mixture \cite{nakaharap}.
  From this measurement, they demonstrated that the mixture memorizes
  an external force before drying
  only when it behaves as a viscoplastic material with a finite yield stress.
  Moreover, they report that 
  only when the strength of the force is larger than a threshold,
  the mixture memorizes the force.
  The relation between the memory of the force 
  and a finite yield stress supports our conjecture 
  that a plastic deformation plays an important role in the 
  memory effect on the formation of drying cracks.
  Furthermore, 
  the existence of the threshold coincides with the result of our model.

\section*{Acknowledgments}
We thank A. Nakahara and Y. Matsuo for showing us experiments 
and for stimulating discussions. We also thank S. Sasa, T. S. Komatsu,
and T. Ooshida for their valuable discussions.

\appendix
\section{THE ANALYTICAL SOLUTION OF THE POSITION OF THE FIRST CRACK}
\label{sol}

  In this appendix, we show the path to obtain Eqs. (\ref{xc0}) and (\ref{xc}),
  which determine the position of the first perpendicular crack $x_c$.
  This appendix consists of two subsections.
  In the first subsection,
  we calculate $S(x,T_1)$,
  and in the second subsection, we calculate $T_{xx}(x,t)$ ($t>T_3$).
  Substituting the expression of $T_{xx}(x,t)$ 
  into Eqs. (\ref{txc}) and (\ref{dtxc}),
  we obtain Eqs. (\ref{xc0}) and (\ref{xc}).
  In these calculations, we assume that $T_1<T_2<T_3$ 
  in order to simplify the calculations.

  \subsection{The calculation of $S(x,T_1)$}

    First, we evaluate the minimum external force $\alpha_{Y0}$, 
    which yields a plastic deformation.
    Assuming that a plastic deformation does not occur ($S(x,t)=0$)
    and solving Eq. (\ref{du}) 
    under the boundary condition (\ref{dbbc}),
    we obtain 
    \begin{eqnarray}
    U(x,t) = \frac{\alpha (t) H^2}{\mu}
    \left(1-\frac{\cosh qx}{\cosh qL}\right), \label{us0}
    \end{eqnarray}
    where 
    \begin{eqnarray}
    q^2 = \frac{\mu}{(\lambda + 2\mu)H^2}.
    \end{eqnarray}
    By substituting Eq. (\ref{us0}) into Eq. (\ref{dsxz}),
    we obtain a shear stress $T_{xz}(x,t)$ as
    \begin{eqnarray}
    T_{xz}(x,t)=\alpha (t) H \left(1-\frac{\cosh qx}{\cosh qL}\right).
    \end{eqnarray}
    The assumption that $S(x,t)=0$ is valid
    if the equation
    \begin{eqnarray}
    T_{xz}(x,t) < \sigma_{Y0}
    \end{eqnarray}
    is satisfied.
    This condition is equivalent to 
    \begin{eqnarray}
    \alpha (t) < \alpha_{Y0},
    \end{eqnarray}
    where
    \begin{eqnarray}
    \alpha_{Y0}=\frac{\sigma_{Y0} \cosh qL}{H(\cosh qL -1)}. \label{ay0}
    \end{eqnarray}
    Hence, we find that if $\alpha_M<\alpha_{Y0}$,
    \begin{eqnarray}
    S(x,T_1)=0. \label{sx0}
    \end{eqnarray}

    If $\alpha_M>\alpha_{Y0}$,
    a plastic deformation occurs.
    Then, 
    assuming that a plastic deformation occurs in the region $-x_s<x<x_s$,
    we calculate $S(x,T_1/2)$. 
    From Eq. (\ref{S}),
    $S(x,t)$ is given by 
    \begin{eqnarray}
	 S(x,t) =
	\left \{
	\begin{array}{cc}
	  0 & x < -x_s \\
	  \displaystyle \frac{U}{H}-\frac{\sigma_{Y0}}{\mu} & -x_s < x < x_s \\
	  0 & x_s < x.
	\end{array}
	\right. 
	\label{st}
    \end{eqnarray}
    By substituting this into Eq. (\ref{du}) and noting 
    $\alpha(T_1/2)=\alpha_M$,
    we obtain the equations of $U(x,T_1/2)$ as
    \begin{eqnarray}
	\left \{
	\begin{array}{cc}
	  \displaystyle (\lambda + 2\mu)\frac{\partial^2 U}{\partial x^2} 
	  -\frac{\mu}{H^2}U + \alpha_M =0 & x < -x_s \\
	  \displaystyle (\lambda + 2\mu)\frac{\partial^2 U}{\partial x^2} 
	  -\frac{\sigma_{Y0}}{H} + \alpha_M =0 & -x_s < x < x_s \\
	  \displaystyle (\lambda + 2\mu)\frac{\partial^2 U}{\partial x^2} 
	  -\frac{\mu}{H^2}U + \alpha_M =0 & x_s < x.
	\end{array}
	\right. 
    \end{eqnarray}
    By solving these equations under the boundary conditions (\ref{dbbc})
    and under the matching conditions that
    $U(x,T_1/2)$, $T_{xx}(x,T_1/2)$, and $S(x,T_1/2)$ 
    are continuous at $x=\pm x_s$, 
    we obtain the expression of $U(x,T_1/2)$ 
    and the equation to determine $x_s$.
    The expression of $U(x,T_1/2)$ is
     \begin{widetext}
     \begin{eqnarray}
	 U(x,T_1/2) =
	\left \{
	\begin{array}{cc}
	  \displaystyle \frac{\alpha_M H^2}{\mu} +\frac{A_s H}{\mu}B(x,-x_s)
	   & x < -x_s \\
	  \displaystyle \frac{A_s H}{2\mu} q^2(x^2-x_s^2) 
	  + \frac{\sigma_{Y0}H}{\mu}& -x_s < x < x_s  \\
	  \displaystyle \frac{\alpha_M H^2}{\mu} + \frac{A_s H}{\mu}
	  B(x,x_s) & x_s < x, 
	\end{array}
	\right. 
	\label{ut12}
    \end{eqnarray}
     \end{widetext}
    where 
    \begin{eqnarray}
    A_s = \sigma_{Y0}-\alpha_M H
    \end{eqnarray}
    and
    \begin{equation}
    B(x_1,x_2) = \cosh q(x_1-x_2) + qx_2 \sinh(x_1-x_2). 
    \end{equation}
    The equation to determine $x_s$ is 
    \begin{eqnarray}
    \alpha_M H + A_s B(L,x_s)= 0.
    \end{eqnarray}
    Substituting Eq. (\ref{ut12}) into Eq. (\ref{st}),
    we obtain $S(x,T_1/2)$ as
     \begin{eqnarray}
	 S(x,T_1/2) =
	\left \{
	\begin{array}{cc}
	  0 & x < -x_s \\
	  \displaystyle \frac{A_s}{2\mu} q^2(x^2-x_s^2) & -x_s < x < x_s \\
	  0 & x_s < x.
	\end{array}
	\right. 
	\label{st1}
    \end{eqnarray}

    Finally, $S(x,T_1)$ is calculated.
    First, we calculate $U(x,T_1)$
    on the assumption that 
    \begin{eqnarray}
    S(x,T_1) = S(x,T_1/2).\label{st2}
    \end{eqnarray}
    Substituting Eq. (\ref{st2}) into Eq. (\ref{du}) and noting $\alpha(t)=0$,
    we obtain the equation of $U(x,T_1)$ as
    \begin{widetext}
    \begin{eqnarray}
	\left \{
	\begin{array}{cc}
	  \displaystyle (\lambda + 2\mu)\frac{\partial^2 U}{\partial x^2} 
	  -\frac{\mu}{H^2}U  =0 & x < -x_s \\
	  \displaystyle (\lambda + 2\mu)\frac{\partial^2 U}{\partial x^2} 
	  -\frac{\mu}{H^2}U +\frac{A_s}{2H} q^2(x^2-x_s^2)=0
	   & -x_s < x < x_s \\
	  \displaystyle (\lambda + 2\mu)\frac{\partial^2 U}{\partial x^2} 
	  -\frac{\mu}{H^2}U =0 & x_s < x.
	\end{array}
	\right. 
	\label{a0u}
    \end{eqnarray}
    \end{widetext}
    By solving this equation under the boundary conditions and 
    the matching conditions,
    we obtain $U(x,T_1)$ as
    \begin{widetext}
     \begin{eqnarray}
	 U(x,T_1) =
	\left \{
	\begin{array}{cc}
	  \displaystyle \frac{A_s H}{\mu}\left(B(x,-x_s)-B(L,x_s)
	  \frac{\cosh qx}{\cosh qL}\right)
	   & x < -x_s \\
	  \displaystyle \frac{A_s H}{\mu} \left(1+\frac{q^2(x^2-x_s^2)}{2} 
	  -B(L,x_s) \frac{\cosh qx}{\cosh qL}\right)& -x_s < x < x_s  \\
	  \displaystyle \frac{A_s H}{\mu}\left(B(x,x_s)-B(L,x_s)
	  \frac{\cosh qx}{\cosh qL}\right)
	  & x_s < x. 
	\end{array}
	\right. 
    \end{eqnarray}
    \end{widetext}
    Substituting this equation and Eq. (\ref{st2}) into Eq. (\ref{dsxz}),
    we get $T_{xz}(x,T_1)$ as
    \begin{widetext}
     \begin{eqnarray}
	 T_{xz}(x,T_1) =
	\left \{
	\begin{array}{cc}
	  \displaystyle A_s\left(B(x,-x_s)-B(L,x_s)
	  \frac{\cosh qx}{\cosh qL}\right)
	   & x < -x_s \\
	  \displaystyle A_s \left(1 
	  -B(L,x_s) \frac{\cosh qx}{\cosh qL}\right)& -x_s < x < x_s  \\
	  \displaystyle A_s\left(B(x,x_s)-B(L,x_s)
	  \frac{\cosh qx}{\cosh qL}\right)
	  & x_s < x. 
	\end{array}
	\right. 
    \end{eqnarray}
    \end{widetext}
    The assumption that $S(x,T_1) = S(x,T_1/2)$ is valid
    when the equation
    \begin{eqnarray}
    |T_{xz}(x,T_1)| < \sigma_{Y0}
    \end{eqnarray}
    is satisfied,
    because $S(x,T_1) = S(x,T_1/2)$ implies that 
    a plastic deformation does not occur for $T_1/2 < t <T_1$.
    This condition is equivalent to 
    \begin{eqnarray}
    \alpha_M < \alpha_{Y1},
    \end{eqnarray}
    where
    \begin{eqnarray}
    \alpha_{Y1}=\frac{\sigma_{Y0} (2\cosh qL - B(L,x_s))}{\cosh qL -B(L,x_s)}.
    \end{eqnarray}
    Hence, if 
    \begin{eqnarray}
    \alpha_{Y0} < \alpha_M < \alpha_{Y1},
    \end{eqnarray}
    then $S(x,T_1)$ is expressed by Eq. (\ref{st2}).

    \subsection{The calculation of $T_{xx}(x,t)$ for $t>T_3$}

    In this subsection, we calculate $T_{xx}(x,t)$ for $t>T_3$
    in the cases $\alpha_M < \alpha_{Y0}$ 
    and $\alpha_{Y0} < \alpha_M < \alpha_{Y1}$.

    If $\alpha_M < \alpha_{Y0}$,
    then $S(x,T_1)=0$.
    Since $\sigma_Y=\infty$,
    $S(x,t)=0$ for $t>T_3$.
    Substituting $S(x,t)=0$ and $\alpha(t)=0$ into Eq. (\ref{du}),
    we obtain the equation of $U(x,t)$ for $t>T_3$ as
    \begin{eqnarray}
    (\lambda + 2\mu)\frac{\partial^2 U}{\partial x^2} 
    -\frac{\mu}{H^2}U =0.
    \end{eqnarray}
    By solving this equation under the boundary conditions (\ref{dabc}),
    we obtain $U(x,t)$ for $t>T_3$ as
    \begin{eqnarray}
    U(x,t)=-\frac{c(t) \sinh qx}{q \cosh qL}.
    \end{eqnarray}
    Substituting this into Eq. (\ref{dsxx}),
    we get $T_{xx}(x,t)$ in the case $\alpha_M < \alpha_{Y0}$ as
    \begin{eqnarray}
    T_{xx}(x,t)=(\lambda+2\mu)c(t)\left(1-\frac{ \cosh qx}{ \cosh qL}\right).
    \end{eqnarray}

    If $\alpha_{Y0} < \alpha_M < \alpha_{Y1}$,
    $S(x,T_1)$ is expressed by Eq. (\ref{st2}).
    Since $\sigma_Y=\infty$,
    \begin{eqnarray}
    S(x,t)=S(x,T_1) \label{sxt3}
    \end{eqnarray}
    for $t>T_3$,
    where $S(x,T_1)$ is expressed by Eq. (\ref{st2}).
    Substituting Eq. (\ref{sxt3}) into Eq. (\ref{du}) and noting $\alpha(t)=0$,
    we obtain the same equation of $U(x,t)$ as Eq. (\ref{a0u}). 
    By solving this equation under the boundary conditions
    (\ref{dabc}) and the matching conditions,
    we obtain $U(x,t)$ for $t>T_3$ as
    \begin{widetext}
    \begin{eqnarray}
	 U(x,t) =
	\left \{
	\begin{array}{lc}
	  \displaystyle -\frac{c(t) \sinh qx}{q \cosh qL} 
	   +\frac{A_s H}{\mu}
	  \left(B(x,-x_s)-D(L,x_s)\frac{\cosh qx}{\sinh qL} \right)
	   & x < -x_s \\
	  \displaystyle -\frac{c(t) \sinh qx}{q \cosh qL} 
	   +\frac{A_s H}{\mu} \left(1+\frac{q^2(x^2-x_s^2)}{2} 
	  -D(L,x_s) \frac{\cosh qx}{\sinh qL}\right)& -x_s < x < x_s  \\
	  \displaystyle -\frac{c(t) \sinh qx}{q \cosh qL} 
	  +\frac{A_s H}{\mu}
	  \left(B(x,x_s)-D(L,x_s)\frac{\cosh qx}{\sinh qL} \right)
	  & x_s < x, 
	\end{array}
	\right. 
    \end{eqnarray}
    \end{widetext}
    where 
    \begin{equation}
    D(x_1,x_2)  =  \sinh q(x_1-x_2) + q x_2\cosh q(x_1-x_2).
    \end{equation}
    Substituting this into Eq. (\ref{dsxx}),
    we get $T_{xx}(x,t)$ for $t>T_3$ 
    in the case $\alpha_{Y0} < \alpha_M < \alpha_{Y1}$ as
    \begin{widetext}
    \begin{eqnarray}
	 T_{xx}(x,t) =
	\left \{
	\begin{array}{lc}
	  \displaystyle (\lambda+2\mu)c(t)\left(1-\frac{ \cosh qx}{ \cosh qL}\right) 
	  +\frac{A_s}{qH}
	  \left(D(x,-x_s)-D(L,x_s)\frac{\sinh qx}{\sinh qL} \right)
	   & x < -x_s \\
	  \displaystyle (\lambda+2\mu)c(t)\left(1-\frac{ \cosh qx}{ \cosh qL}\right) 
	  +\frac{A_s}{qH} \left(qx 
	  -D(L,x_s) \frac{\sinh qx}{\sinh qL}\right)& -x_s < x < x_s  \\
	  \displaystyle (\lambda+2\mu)c(t)\left(1-\frac{ \cosh qx}{ \cosh qL}\right)
	  +\frac{A_s}{qH}
	  \left(D(x,x_s)-D(L,x_s)\frac{\sinh qx}{\sinh qL} \right)
	  & x_s < x. 
	\end{array}
	\right. 
    \end{eqnarray}
    \end{widetext}


\begin{thebibliography}{10}

\bibitem{groisman} G. Groisman and E. Kaplan, Europhys. Lett {\bf 25}, 415 (1994).
\bibitem{nakahara} A. Nakahara and Y. Matsuo, Bussei Kenkyuu {\bf 81}, 2 (2003).
\bibitem{sand} L. Vanel, D. Howell, D. Clark, R. P. Behringer and E. Clement, Phys. Rev. E {\bf 60}, R5040 (1999).
\bibitem{paste} M. Cloitre, R. Borrega and L. Leibler, Phys. Rev. Lett. {\bf 85},  4819 (2000).
\bibitem{rubber} Y. Miyamoto, K. Fukao, H. Yamao and K. Sekimoto, Phys. Rev. Lett. {\bf 88}, 225504 (2002).
\bibitem{waribasi} T. Ooshida and K. Sekimoto, cond-mat/0410306. 
\bibitem{kitsune} S. Kitsunezaki, Phys. Rev. E {\bf 60}, 6449 (1999).
\bibitem{sasa} T. S. Komatsu and S. Sasa, Jpn. J. Appl. Phys. {\bf 36}, 391 (1997).
\bibitem{nakaharap} A. Nakahara and Y. Matsuo, J. Phys. Soc. Jpn. {\bf 74}, 1362 (2005).

\end{thebibliography}
\end{document}